\documentclass[aps,prl,twocolumn,10pt]{revtex4-2}
\usepackage{amsfonts,amssymb,amsmath}            
\newcommand{\ket}[1]{\left | #1 \right\rangle}
\newcommand{\bra}[1]{\left \langle #1 \right |}
\newcommand{\Tr}{\text{Tr}}

\newcommand{\proj}[1]{\ket{#1}\bra{#1}}
\DeclareFontFamily{U}{bbold}{}
\DeclareFontShape{U}{bbold}{m}{n}
 {
  <-5.5> s*[1.069] bbold5
  <5.5-6.5> s*[1.069] bbold6
  <6.5-7.5> s*[1.069] bbold7
  <7.5-8.5> s*[1.069] bbold8
  <8.5-9.5> s*[1.069] bbold9
  <9.5-11> s*[1.069] bbold10
  <11-15> s*[1.069] bbold12
  <15-> s*[1.069] bbold17
 }{}

\DeclareRobustCommand{\identity}{%
  \text{\usefont{U}{bbold}{m}{n}1}%
}

\begin{document}
\title{A Note on the Speed of Perfect State Transfer}
\author{Alastair Kay}
\affiliation{Department of Mathematics, Royal Holloway University of London, Egham, Surrey, TW20 0EX, UK}
\email{alastair.kay@rhul.ac.uk}
\author{Weichen Xie}\affiliation{Department of Mathematics, Clarkson University, Potsdam, New York, USA 13699-5815.}\email{xiew@clarkson.edu}
\author{Christino Tamon}
\affiliation{Department of Computer Science, Clarkson University, Potsdam, New York, USA 13699-5815.}
\email{tino@clarkson.edu}
\date{\today}
\begin{abstract}
In Phys.\ Rev.\ A 74, 030303 (2006), Yung showed that for a one-dimensional spin chain of length $N$ and maximum coupling strength $J_{\max}$, the time $t_0$ for a quantum state to transfer from one end of the chain to another is bounded by $J_{\max} t_0\geq\pi N/4$ (even $N$) and $J_{\max} t_0\geq\pi\sqrt{N^2-1}/4$ (odd $N$). The proof for even $N$ was elegant, but the proof for odd $N$ was less so. This note provides a proof for the odd $N$ case that is simpler, and more in keeping with the proof for the even case.
\end{abstract}
\maketitle

The benefit of this simplified proof is that it can be used elsewhere. For example, in the study of speed limits for synthesising quantum states using identical Hamiltonian structures \cite{kay2017}, where the phase conditions on the eigenvalues are slightly different, these differences are readily incorporated in the optimisation. Studies of fractional revival \cite{genest2016} are another option.

A tridiagonal $N\times N$ matrix with real diagonal elements $B_n$ and positive off-diagonal elements $J_n$,
$$
H=\sum_{n=1}^NB_n\proj{n}+\sum_{n=1}^{N-1}J_n\left(\ket{n}\bra{n+1}+\ket{n+1}\bra{n}\right),
$$
is capable of perfect transfer ($e^{-iHt_0}\ket{1}=e^{i\phi}\ket{N}$) if and only if $H$ is symmetric, meaning $B_n=B_{N+1-n}$ and $J_n=J_{N-n}$, and the ordered eigenvalues $\lambda_n>\lambda_{n+1}$ satisfy $e^{-i\lambda_nt_0}=(-1)^{n+1}e^{i\phi}$ for some real parameters $t_0$ (the transfer time) and $\phi$ \cite{kay2010a}. The symmetry imposes that the eigenvectors satisfy $S\ket{\lambda_n}=(-1)^{n+1}\ket{\lambda_n}$ where
$$
S=\sum_{n=1}^N\ket{n}\bra{N+1-n}.
$$

{\em N even:} We repeat Yung's proof \cite{yung2006}. If we calculate $\Tr(SH)$ for both the matrix directly, and for the eigenvector decomposition, we have
$$
2J_{N/2}=\sum_{n=1}^{N/2}\lambda_{2n-1}-\lambda_{2n}.
$$
We know that $\lambda_{2n-1}\geq \lambda_{2n}+\frac{\pi}{t_0}$ in order for it to satisfy the required phase property for the eigenvalues, so
$$
2J_{\max}\geq2J_{N/2}\geq\frac{N}{2}\frac{\pi}{t_0},
$$
immediately yielding the claimed relation.

{\em N odd:} This more involved case starts similarly,
$$
B_{(N+1)/2}=\Tr(HS)=\sum_n\lambda_{2n-1}-\sum_n\lambda_{2n}.
$$

Next, we evaluate $\Tr(SH^2)$
$$
\Tr(SH^2)=4J_{(N-1)/2}^2+B_{(N+1)/2}^2=\sum_n\lambda_{2n-1}^2-\sum_n\lambda_{2n}^2.
$$
Let the vector $\underline{\lambda}=(\lambda_1,\lambda_3,\ldots,\lambda_N,\lambda_2,\dots\lambda_{N-1})$. Then we can express
$$
4J_{(N-1)/2}^2=\underline{\lambda}^T\left(\begin{array}{cc}
\identity_{(N+1)/2}-\textbf{1} & \textbf{1} \\ \textbf{1} & -\identity_{(N-1)/2}-\textbf{1}
\end{array}\right)\underline{\lambda},
$$
where $\textbf{1}$ is the all-ones matrix of the appropriate size, sizes of the blocks being specified by the subscript of the $\identity$. However, there are many equivalent matrices (since, for example, $\lambda_1\lambda_2$ appears in two terms). One such is
$$
\tilde A=2\left(\begin{array}{cc}
-U & \identity+U \\ U & -\identity-U
\end{array}\right)
$$
where $U$ is the upper triangular component of $\textbf{1}=\identity+U+U^T$ (of the appropriate size) -- just evaluate $\frac12(\tilde A+\tilde A^T)$. Note that rows for terms $\lambda_{2n-1}$ and $\lambda_{2n}$ are identical up to a sign, as are columns for $\lambda_{2m}$ and $\lambda_{2m+1}$. Hence
$$
4J_{(N-1)/2}^2=2\sum_{n=1}^{(N-1)/2}(\lambda_{2n-1}-\lambda_{2n})\sum_{m=n}^{(N-1)/2}(\lambda_{2m}-\lambda_{2m+1}).
$$
Since $\lambda_n-\lambda_{n+1}\geq \frac{\pi}{t_0}$, we bound this by
$$
4J_{(N-1)/2}^2\geq 2\frac{\pi^2}{t_0^2}\sum_{n=1}^{(N-1)/2}\left(\frac{N+1}{2}-n\right)=\frac{\pi^2}{t_0^2}\frac{N^2-1}{4}.
$$
This leaves the claimed result
$$
J_{\max}t_0\geq\frac{\pi\sqrt{N^2-1}}{4}.
$$

{\em Acknowledgements}: AK was supported by EPSRC grant EP/N035097/1. 
%

\end{document}